\documentclass[mathleft]{an}
\pdfoutput=1   
\usepackage{graphicx}
\usepackage{times}
\usepackage{natbib}
\bibpunct{(}{)}{;}{a}{}{,}
\usepackage{color}
\usepackage{hyperref}
\hypersetup{colorlinks=true,urlcolor=black,linkcolor=red,citecolor=blue}

\begin{document}

\Pagespan{1}{6}
\Yearpublication{2014}%
\Yearsubmission{2014}%
\Month{00}%
\Volume{000}%
\Issue{00}%

\DOI{10.1002/asna.201412061}%

\title{
Positive and negative feedback by AGN jets in high-redshift galaxies 
}

\author{V. Gaibler\thanks{Corresponding author:
  \email{v.gaibler@uni-heidelberg.de}\newline}
}
\titlerunning{Positive and negative AGN feedback}
\authorrunning{V. Gaibler}
\institute{
Universit\"at Heidelberg, Zentrum f\"ur Astronomie, Institut f\"ur Theoretische Astrophysik,\\
Albert-Ueberle-Str. 2, 69120 Heidelberg, Germany\\
v.gaibler@uni-heidelberg.de}

\received{2014 Apr 7}
\accepted{2014 Apr 10}
\publonline{later}


\abstract{%
Simulations of feedback by jets from active galactic nuclei (AGN) in the past mostly focused on the interaction at large scales as the circumgalactic medium or intra-cluster medium for clusters of galaxies. Only in recent years, simulations have included the interaction of jets with a highly inhomogeneous medium as required by a multi-phase interstellar medium (ISM). At the same time, feedback by AGN has become a common component for cosmological simulations of galaxy evolution to form massive galaxies compatible with observations. I will present some of our recent results and will put them into further context of other feedback simulations and how the opposing effects of positive and negative feedback by jets might be understood in terms of different properties of the ISM.
}

\maketitle


\section{Introduction}

Feedback by active galactic nuclei (AGN) has become a commonly invoked process in galaxy evolution. It acts on two scales: large-scale feedback on the circumgalactic gas and (for clusters of galaxies) the intra-cluster medium on the one hand, the galaxy-scale feedback on the interstellar medium (ISM) of the galaxy on the other. For the large scales, there are now great observations of cavities in the hot X-ray gas \citep{McNamaraNulsen2007} that can be used as calorimeters of the jet power by linking their volumes to the environmental pressure. Since these cavities are found to be filled by radio-emitting jet plasma, this is a quite robust approach that showed that on these scales, feedback may be sufficient to offset the gas cooling and explain the lack of strong cooling flows in clusters of galaxies. Furthermore it allows much better determination of time-averaged jet powers than possible through its radio emission.

On galactic scales, however, AGN feedback still has to be considered a mostly theoretical construct. While there are detailed observations available that show interaction of the AGN with interstellar gas, both for jets and winds, it is far from clear what their actual importance is on the thermal budget and evolution of the host galaxy. On the theoretical side, AGN feedback has been deemed responsible for a number of observed findings, the most prominent ones being the suppression or quenching of star formation in massive galaxies at late times (negative feedback) and the correlation of the central black hole masses with the galactic spheroid mass or velocity dispersion. This is a reasonable approach since AGN activity is a frequent phenomenon in massive galaxies \citep[e.g.][for radio-loud AGN]{Best+2005} and the corresponding energies can be very large (both jet power from cavities and radiative power from quasars). Earlier cosmological hydrodynamical simulations failed to produce realistic galaxies at the 
high-mass end since they continued to grow also at low redshifts and are still dominated by star formation while observed massive galaxies are mostly elliptical galaxies with only little ongoing star formation. Semianalytic models \citep[e.g.][]{Croton+2006,Bower+2006} showed that introducing AGN feedback can bring massive galaxies to agreement with observations, and this could also be found in more recent cosmological simulations that included AGN feedback \citep[e.g.][]{Sijacki+2007,Dubois+2010,McCarthy+2010,Vogelsberger+2013}. It should be mentioned that these simulations generally do not make a distinction between feedback at the two scales that are closely linked in cosmological simulations. Yet, feedback at large scales is qualitatively different from galaxy scales in that it involves almost exclusively the interaction of the AGN with a hot and low density gas. On galaxy scales, in contrast, the feedback occurs in the interstellar medium (ISM) that has gas densities varying over several orders of magnitude and a fractal, cloudy structure. For most simulations, these dense stuctures are quite below the resolution limit.

Although the model of AGN feedback is largely successful, it is important to recall that the underlying feedback process is still not well understood. Only recently hydrodynamic simulations of AGN jets have included a largely inhomogeneous environment and the impact of the jet on the gas dynamics and star formation in the ISM is studied in more detail. I will describe some of the findings in the following and discuss issues that will have to be addressed in the future to improve our understanding of this feedback and eventually allow a better modelling of AGN feedback in galaxy evolution and rigorous comparison to observations.

One should keep in mind that not all observed relations used in favour of AGN feedback do actually underpin its importance. While several feedback models provide explanations for the observed black hole scaling relations between the black hole and the host galaxy \citep[e.g.][for the $M$--$\sigma$ relation]{SilkRees1998,King2003}, \citet{JahnkeMaccio2011} argue that the similar scaling between black hole mass and the galaxy bulge mass \citep{Magorrian+1998} could simply be due to a statical convergence process and does not require AGN feedback.

\section{Types of AGN feedback}

Feedback by nuclear supermassive black holes includes various phenomena and a rich nomenclature. However, it can be categorized into two main classes: accretion-disk-driven feedback and jet-driven feedback, often referred to as ``quasar mode'' and ``radio mode'' feedback. The former feedback stems from the accretion onto the black hole in a geometrically thin but optically thick accretion disk or for weaker accretion a geometrically thick but optically thin radiatively inefficient accretion flow. The corresponding action on the environment is ultimately driven thermally or by radiation. Coupling of the radiation to gas and dust can result in mechanical outflows or winds with large opening angles. The other class, jet-driven feedback, originates from the formation of collimated, relativistic jets from the innermost accretion flow either by magnetocentrifugal launching from the disk or the by the black hole spin. This type of feedback is driven by the momentum of the narrow, supersonic jet beam (initially Poynting flux dominated jets would eventually become matter-dominated as the jets expand). The jets and jet-inflated cocoons/lobes are generally observable by their radio synchrotron emission, but also at other energies as optical and X-rays.

Clearly, jets are also linked to accretion onto the black hole, but there is no simple and direct connection between those two processes since jet formation has additional requirements, e.g. for the magnetic field. Only a minority of strongly accreting black holes produces powerful jets. Furthermore, if jets can be powered by the black hole spin, the black holes may store large amounts of energy from accreted matter in their rotational energy and release it much later when the accretion rates are considerably different. This might essentially decouple the jet powers from the current accretion rates and allow more degrees of freedom for jet feedback than for accretion-driven feedback.

Although it can be expected that in real objects both types of feedback often will be operating simulaneously, though on potentially very different levels, separating these types for basic theoretical studies is helpful to understand in more detail the important physical processes. Since outflows can be produced by both feedback types, it is also important to consider what parameters enter the models and simulations and which observed class objects they correspond to. In particular, jets and winds are sometimes not clearly separated in theoretical studies and hence may lead to contradictions with observations. In the following, I will focus more on the feedback from jets than from radiation-driven feedback.

Observations of jets on pc scales with VLBI show apparent superluminal speeds and strong brightness asymmetries between the two jet sides due to Doppler beaming, often with only one jet visible (but the large scale still shows the lobes from the other side). This is generally regarded as evidence for highly relativistic speeds of the jet beam with Lorentz factors of $\sim 10$. On the kpc scale, if jets don't decollimate and form FR I radio sources \citep{FanaroffRiley1974}, they still remain mildly relativistic with typical speeds in the range of $0.5$--$0.8\,c$ \citep{MullinHardcastle2009}, although X-ray observations may require the presence of a jet spine that is faster than that. While the jet density is not directly measurable from the synchrotron brightness, it can be derived from the kinetic jet powers (cavity powers), speeds and the observable jet radius, finding that it is $\leq 10^{-4} \: m_\mathrm{p} \, \mathrm{cm}^{-3}$. Other estimates or limits can be obtained from the small mass flux channelled to the jet relative to the Eddington accretion rate, the jet head propagation speed (which is much smaller than the jet beam speed) or the hotspot termination pressures determined from the synchrotron emission. To describe the feedback of actual jets with the galactic gas, it is hence important to use jet properties compatible with observed jets. The large jet speeds have the unfortunate consequence for numerical simulations to limit time steps to very small values which is computationally very expensive but is necessary to model realistic radio sources -- for large-scale cosmological simulations this may even be prohibitive currently. Subrelativistic, denser outflows, in contrast, are rather linked with AGN winds than with jets and the ``radio mode''.

\section{AGN feedback in a multi-phase ISM}

\subsection{Positive and negative jet feedback}

\begin{figure*}
\centering
\includegraphics[width=0.8\textwidth]{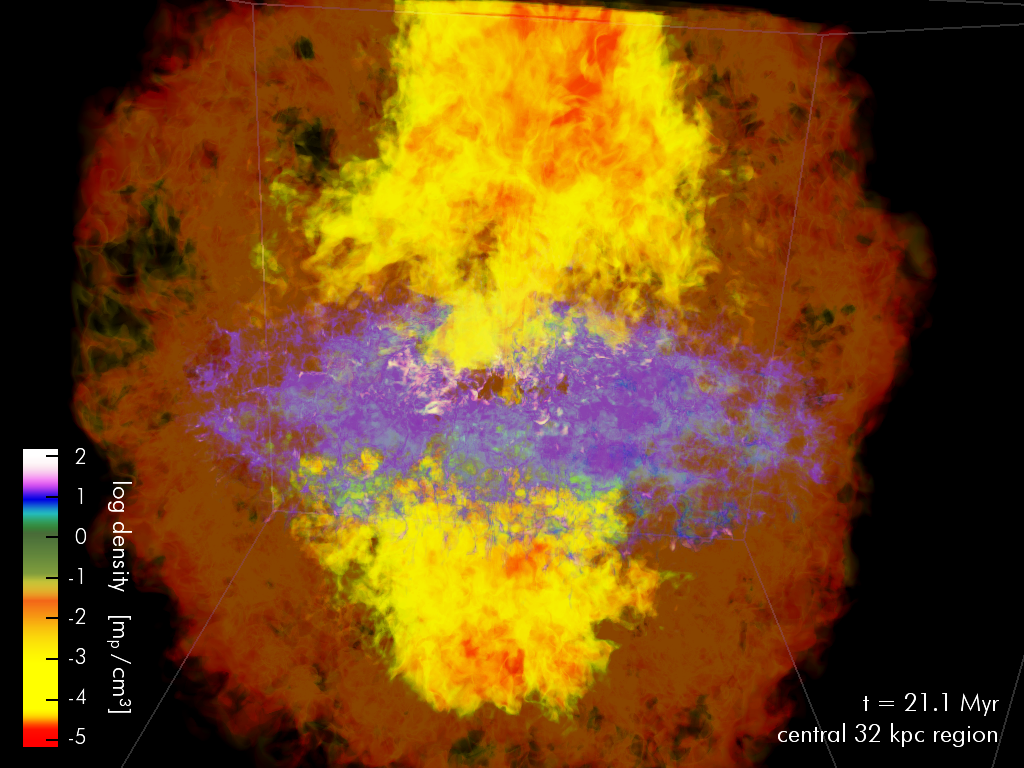} 
\caption{Volume rendering of the jet and the gas disk after is has broken out vertically, derived from the simulation of \citet{Gaibler+2012}. The colors correspond to the densities indicated by the colorbar, the low-density circumgalactic gas is rendered transparently. While the central part of the disk has been cleared by the blastwave, the outer parts of the disk are mostly unaffected by the negative feedback but get compressed by the surrounding overpressured jet cocoon of low density.}
\label{densityview}
\end{figure*}
Jets don't propagate in empty space but in or through an interstellar medium (ISM) that is generally much denser than the jet plasma. At the smallest scales, jets are conical and propagate ballistically, not noticing the inertia of the ambient gas. On a scale  $L_\mathrm{1b}$ \citep{Krause+2012}, $\sim$ pc or less for most cases, the expanding jet becomes underdense, its forward thrust is too low for ballistic motion and propagation of the jet termination point, the hotspot, becomes much smaller than the bulk speed of the jet beam. Around this point, the jet starts inflating a cocoon and thermalizes its kinetic energy more efficiently with stronger density contrast. Where exactly this happens depends on the ISM density structure in the nuclear region. The efficient thermalization produces a blastwave originating from this region and while the inner jet was highly directed (collimated to a small opening angle), the working surface suddenly becomes much larger (almost isotropic). Now the feedback becomes energy-driven with a mechanical advantage $\gg 1$ and the outward momentum acting on the ISM is much larger than the momentum carried by the jet on the inner scales. Only once the blastwaves originating from the two ``thermalization points'' break out of the ISM (vertically for a globally disk-shaped ISM), the driving cocoon pressure vents out to the circumgalactic medium, the expansion of the blastwave slows down and the radio source can propagate to  large scales. Since the jet there is still much underdense, its bow shock expands laterally, eventually encloses the entire galaxy, and then pressurizes the ISM of the galaxies from the outside (Fig.~\ref{densityview}). Roughly, one could say that jets are masters of deception: the thin and powerful beam and radio lobes get all the attention, but secondary effects as blastwave formation do most of the work on the galaxy scale.

A crucial ingredient of this scenario is the multi-phase nature of the ISM. Its clumpy and fractal structure is found in observations and generally expected within a supernova-regulated multi-phase medium \citep{McKeeOstriker1977}. The densities span over many orders of magnitude, although the filling factors of the denser gas components are considerably below unity. Most jet simulations in the past assumed an ambient gas with only small density gradients, as suitable for the diffuse gas around galaxies. \citet{SutherlandBicknell2007} newly examined the interaction of jets with a fractal gas distribution and cloud sizes up to 50 pc. They and subsequent studies \citep{Gaibler+2011,WagnerBicknell2011,Wagner+2012} found a qualitatively different evolution from jets in a smooth ambient medium as described in the previous paragraph. \citet{Wagner+2012} used a cloud setup that contains a large number of clouds at high resolution, on average distributed evenly within a $1$ kpc sphere. They find that for jets more powerful than $\sim 10^{-4}$ of the Eddington luminosity, feedback can be effective in ejecting gas out of the galaxy and that the efficiency depends much more on the cloud sizes than on the filling factor; clouds smaller than $\sim 10$ pc probably would be destroyed by ablation. The efficient dispersal and ejection of clouds as sites of star formation would hence lead to negative feedback.

With a different setup, \citet{Gaibler+2011} examine the jet feedback to larger scales and times, but at lower resolution. Here, an entire thick and clumpy galactic gas disk is simulated without an upper limit on cloud sizes. Accordingly the disk is more clumpy on large scales and the jet propagation through the ISM is considerably asymmetric. While the innermost region of the disk is mostly cleared of gas, only little is ejected to large distances. Gas just around this central cavity is compressed strongly due to the expanding blastwave and efficient cooling in the dense gas. The probability distribution function of gas density strongly increases at high densities and ablation and ejection do not seem to be very effective. Comparing to the results of \citet{Wagner+2012}, this may simply be due to the larger cloud sizes. An increase in high density gas mass is also seen in their simulations.

The actual impact on galaxy evolution, however, depends on the changes found for star formation. To examine that, \citet{Gaibler+2012} included a star formation model that is often used in galaxy-scale simulations -- gas above a certain density theshold is converted to stars at a certain efficiency per local free-fall time. While the study finds a strong drop in star formation for the central cleared-out region, the star formation in the surrounding compressed gas increases strongly, and at later times also increases considerably in the outer parts of the disk since it is pressurized by the jet cocoon that then surrounds the entire galaxy. Globally, the star formation rate of the galaxy increases by a factor $>3$ within the simulated $10$--$15$ Myr and does not yet show a trend to decline. \citet{Silk2005} argued that this positive feedback by jets may cause ultraluminous starbursts at high redshift that rapidly build up the massive spheroids.

The fate of the cloud eventually depends on the effectiveness of ablation and fragmentation of the cloud. For 100-pc clouds, \citet{Gaibler+2012} argue that both the cloud crushing times as well as the Kelvin--Helmholtz time scale are still large enough to let them survive. Efficient cooling, magnetic fields in the dense gas and self-gravity (at least for the denser regions inside) further protect the cloud and let the compressive effect dominate and trigger or increase the star formation even if it fragments \citep{Mellema+2002}.

\subsection{Low vs. high redshift}

The ISM of high-redshift ($z \sim 2$--$3$) massive galaxies is rather different from the ISM in the local Universe. While the galaxies still have smaller masses, they have much larger gas fractions and a more clumpy disk morphology due to gravitational instability \citep{Ceverino+2010,Shapley2011}. Star formation seems to occur in much larger regions than the molecular cloud complexes in the local Universe. Furthermore, these galaxies are still actively accreting hot and cold gas that continuously replenishes the galactic gas reservoir for star formation, while in the local Universe, massive galaxies are dominated by ellipticals with only little mass in gas and a largely dominating hot gas component. This quite different scenery naturally changes the interaction of jets with the ISM. It should be expected that the larger clumps make negative feedback by cloud ejection less efficient and the clouds more stable against ablation. The simulations by \citet{Gaibler+2012} with their large clumps were meant to mimic this 
situation and argue for positive feedback occuring in high-redshift massive galaxies. The results of \citet{Wagner+2012} suggest that there may be transition from negative feedback to positive feedback for cloud sizes in the range of $10$--$50$ pc which interestingly is similar to the size of local giant molecular cloud complexes. One might hence speculate that in high-redshift galaxies with their larger cloud complexes, feedback could be mostly positive while it is more negative in the local Universe. Another difference is the more disky morphology at high redshift, where ejection of gas is more difficult compared to a more spherical gas distribution, purely by the amount of gas in the jet-affected regions. Gas outside the inner disk region is hard to affect by negative feedback due to the large gas columns and the radially stalling blast wave in the disk, but it can more easily be affected by positive feedback from the overpressured cocoon surrounding the galaxy at later times. The orientation of the jets with respect to the disk plane has no great importance for this consideration: for the extreme case of a jet propagating near the disk plane, rather improbable even for purely random jet orientations, the gas column is very large and much of the cocoon pressure will be lost vertically to the disk once the blast wave radius has reached about $2$ disk scale heights.

Cosmological simulations are considerably more limited in resolution due to their larger computational domains and larger simulated time scales. The large thermalization found by the resolved feedback simulations above would actually support the widespread subgrid modelling with thermal energy injection. Since a clumpy ISM is hardly feasible in cosmological runs except on the largest scales of the galaxies, subgrid models of the ISM have been employed \citep[e.g.][]{SpringelHernquist2003,DuboisTeyssier2008,SchayeDallaVecchia2008} that describe the thermodynamical response of the multi-phase ISM to changes from outside. But for the case of resolved jets interacting with the ISM, such subgrid models are not expected to generally produce the same results as those described, and the same applies to simply introducing jet-driven blast wave of low density and high pressure. However, \citet{Wagner+2012} argue that for the case of small cloud sizes and filling factors, the single-phase subgrid modelling is a good approximation with the resulting negative 
feedback. For larger clouds or high-redshift galaxies, in contrast, the results would not agree. Resolved AGN jet feedback is particularly difficult if not impossible for simulations employing smoothed particle hydrodynamics (SPH) due to the very low densities in the jet beam. There is almost no mass associated with all the kinetic energy input, requiring extremely small particle masses of only few $M_\odot$. For those simulations, the introduction of a jet-driven blastwave may be a viable way if the ISM is clumpy enough.

For positive feedback it is also important to consider whether the triggering of star formation is ``strong'' or ``weak'' triggering \citep{Dale+2007b}, i.e. whether actually more stars are formed due to the jet activity or the stars would form anyway and star formation is only accelerated (and is smaller later). While this so far cannot be determined from the simulations since they model only a short time and address only isolated galaxies, it can be conjectured that at high redshift there might actually be strong triggering of star formation since there is a continuous supply of infalling gas and an accelerated formation of stars does not lead to a shortage of gas in the galaxy. Further examination of this, however, should be done in future studies.

\subsection{Feedback by AGN winds}

Having focused on AGN jets so far, it is important to also consider accretion disk-driven feedback and in particular the differences compared to jet feedback. The winds formed there are commonly observed in AGN as warm absorbers in UV or X-rays, broad absorption line (BAL) quasar, ultra-fast outflows (UFOs) or quasar outflows in emission lines gas. 
They can be launched by coupling of matter (e.g. from the dusty torus) to the radiation field of the accretion disk and also have been studied for many years \citep[e.g.][]{Proga+2000,Ostriker+2010,NayakshinZubovas2012,Wagner+2013}. In contrast to collimated, relativistic jets, these winds have much larger opening angles and though they can be fast, they are generally subrelativistic, with larger mass fluxes, and are momentum-driven. Hence they appear to be very different from jets. Since we have learned, however, that the interaction of jets with the ISM very quickly turns into a scenario where a jet-driven blastwave interacts with the ISM, the differences become considerably smaller. Similar to AGN winds, jets can also affect a very large solid angle seen from the galactic center. In fact, apart from the radio emission closely linked to jets, it may be difficult to distinguish between the two cases. \citet{Wagner+2013} have applied their fractal ISM setup to UFOs and find that the interaction with the ISM 
is similar. 

It is hence desirable to further study the differences between the two mechanisms, also because they may both be acting simultaneously in some sources. Arguments made against jets affecting the ISM based on momentum arguments have to be viewed in the light of the large mechanical advantage that can be achieved with an energy-driven blastwave. And although both types of feedback act on a large opening angle or near-isotropically, it has be be kept in mind that the geometrical unification of AGN types requires the presence of some thick ``dusty torus'', and providing negative feedback isotropically on star formation in a galaxy seems hard without also getting rid of this structure (and thus unification) first. Interestingly, jet feedback could provide a nice way out of this difficulty since the acting two blastwaves originate roughly from the two locations where the jet density drops strongly below the ambient density, which in principle could be outside the dust torus region. But for both types of feedback, their impact on galaxy evolution will strongly depend on their occurrence rates and either the efficiency of coupling matter to the radiation field or the efficiency of launching jets.

\section{Outlook}

Simulations of both quasar and jet-driven feedback in galaxies have allowed a deeper insight into the driving processes in active galaxies. Many answers are still tentative or speculative, but it is a fascinating finding that both negative and positive feedback may result from the same driving phenomenon. And while it is so far unclear whether the desire for negative feedback in cosmological studies can be fulfilled by AGN jet feedback and this has to be examined further in the next years, there are several additional processes not detailed here that may contribute additionally and have to be further explored, such as heat conduction, the impact of magnetic fields in particular in the dense clouds and the effect of cosmic rays from the nucleus or the jet cocoon on the star formation in the clouds.

\acknowledgements
VG acknowledges financial support by the German Research Foundation
(DFG) through Priority Programme SPP 1177 and in part by Sonderforschungsbereich
SFB 881 ``The Milky Way System'' (subproject B4). The author made use of VAPOR for visualization purposes \citep{ClyneRast2005,Clyne+2007}.


\begin{thebibliography}{}
\expandafter\ifx\csname natexlab\endcsname\relax\def\natexlab#1{#1}\fi

\bibitem[Best {et~al.}(2005)]{Best+2005}
Best, P.N. et~al.: 2005, MNRAS 362, 25

\bibitem[Bower {et~al.}(2006)]{Bower+2006}
Bower, R.G. et~al.: 2006, MNRAS 370, 645

\bibitem[{Ceverino {et~al.}(2010)Ceverino, Dekel, \& Bournaud}]{Ceverino+2010}
Ceverino, D., Dekel, A., Bournaud, F.: 2010, MNRAS 404, 2151

\bibitem[{Clyne {et~al.}(2007)Clyne, Mininni, Norton \& Rast}]{Clyne+2007}
Clyne, J., Mininni, P., Norton, A., Rast, M.: 2007, New Journal of Physics, 9, 301

\bibitem[Clyne \& Rast(2005)]{ClyneRast2005}
Clyne, J., Rast, M.: 2005, Proc. SPIE, Vol. 5669, 284

\bibitem[Croton {et~al.}(2006)]{Croton+2006}
Croton, D.J. et~al.: 2006, MNRAS 365, 11

\bibitem[{Dale {et~al.}(2007)Dale, Clark, \& Bonnell}]{Dale+2007b}
Dale, J.E., Clark, P.C., Bonnell, I.A.: 2007, MNRAS 377, 535

\bibitem[Dubois {et~al.}(2010)]{Dubois+2010}
Dubois, Y., Devriendt, J., Slyz, A., Teyssier, R.: 2010, MNRAS 409, 985

\bibitem[{Dubois \& Teyssier(2008)}]{DuboisTeyssier2008}
Dubois, Y., Teyssier, R.: 2008, A\&A 477, 79

\bibitem[{Fanaroff \& Riley(1974)}]{FanaroffRiley1974}
Fanaroff, B.L., Riley, J.M.: 1974, MNRAS 167, 31

\bibitem[{Gaibler {et~al.}(2011)Gaibler, Khochfar, \& Krause}]{Gaibler+2011}
Gaibler, V., Khochfar, S., \& Krause, M.: 2011, MNRAS 411, 155

\bibitem[Gaibler {et~al.}(2012)]{Gaibler+2012}
Gaibler, V., Khochfar, S., Krause, M., Silk, J.: 2012, MNRAS 425, 438

\bibitem[{Jahnke \& Macci{\`{o}}(2011)}]{JahnkeMaccio2011}
Jahnke, K., Macci{\`{o}}, A.V.: 2011, ApJ 734, 92

\bibitem[King(2003)]{King2003}
King, A. 2003, ApJ 596, L27

\bibitem[Krause {et~al.}(2012)]{Krause+2012}
Krause, M., Alexander, P., Riley, J., Hopton, D.: 2012, MNRAS 427, 3196

\bibitem[Magorrian {et~al.}(1998)]{Magorrian+1998}
Magorrian, J. et~al: 1998, AJ 115, 2285

\bibitem[McCarthy et~al.(2010)]{McCarthy+2010}
McCarthy, I.G. et~al.: 2010, MNRAS 406, 822

\bibitem[{McKee \& Ostriker(1977)}]{McKeeOstriker1977}
McKee, C.F., Ostriker, J.P.: 1977, ApJ 218, 148

\bibitem[{McNamara \& Nulsen(2007)}]{McNamaraNulsen2007}
McNamara, B.R., Nulsen, P.E.J.: 2007, ARA\&A 45, 117

\bibitem[{Mellema {et~al.}(2002)Mellema, Kurk, \& R{\"{o}}ttgering}]{Mellema+2002}
Mellema, G., Kurk, J.D., R{\"{o}}ttgering, H.J.A.: 2002, A\&A 395, L13

\bibitem[{Mullin \& Hardcastle(2009)}]{MullinHardcastle2009}
Mullin, L.M., Hardcastle, M.J. 2009, MNRAS 398, 1989

\bibitem[{Nayakshin \& Zubovas(2012)}]{NayakshinZubovas2012}
Nayakshin, S., Zubovas, K.: 2012, MNRAS 427, 372

\bibitem[Ostriker {et~al.}(2010)]{Ostriker+2010}
Ostriker, J.P. et~al.: 2010, ApJ 722, 642

\bibitem[{Proga {et~al.}(2000)Proga, Stone, \& Kallman}]{Proga+2000}
Proga, D., Stone, J.M., Kallman, T.R.: 2000, ApJ 543, 686

\bibitem[{Schaye \& Dalla~Vecchia(2008)}]{SchayeDallaVecchia2008}
Schaye, J., Dalla~Vecchia, C.: 2008, MNRAS 383, 1210

\bibitem[{Shapley(2011)}]{Shapley2011}
Shapley, A.E.: 2011, ARA\&A 49, 525

\bibitem[Sijacki {et~al.}(2007)]{Sijacki+2007}
Sijacki, D., Springel, V., di~Matteo, T., \& Hernquist, L.: 2007, \mbox{MNRAS} 380,  877

\bibitem[{Silk(2005)}]{Silk2005}
Silk, J.: 2005, MNRAS 364, 1337

\bibitem[{Silk \& Rees(1998)}]{SilkRees1998}
Silk, J., Rees, M.J.: 1998, A\&A 331, L1

\bibitem[{Springel \& Hernquist(2003)}]{SpringelHernquist2003}
Springel, V., Hernquist, L.: 2003, MNRAS 339, 289

\bibitem[{Sutherland \& Bicknell(2007)}]{SutherlandBicknell2007}
Sutherland, R.S., Bicknell, G.V.: 2007, ApJS 173, 37

\bibitem[Vogelsberger {et~al.}(2013)]{Vogelsberger+2013}
Vogelsberger, M. et al.: 2013, MNRAS 436, 3031

\bibitem[{Wagner \& Bicknell(2011)}]{WagnerBicknell2011}
Wagner, A.Y., Bicknell, G.V.: 2011, ApJ 728, 29

\bibitem[{Wagner {et~al.}(2012)Wagner, Bicknell, \& Umemura}]{Wagner+2012}
Wagner, A.Y., Bicknell, G.V., \& Umemura, M.: 2012, ApJ 757, 136

\bibitem[{Wagner {et~al.}(2013)Wagner, Umemura, \& Bicknell}]{Wagner+2013}
Wagner, A.Y., Umemura, M., Bicknell, G.V. 2013, ApJ 763, L18

\end{thebibliography}
\end{document}